\documentclass[12pt, draftcls, onecolumn]{IEEEtran}

\usepackage{color}

\usepackage{bm}

\usepackage{tikz}
\usetikzlibrary{arrows}

\usepackage{url}
\usepackage{amssymb}
\usepackage{dsfont}
\usepackage{amsmath}
\usepackage{xcolor}
\usepackage{array}

\usepackage{algorithm}
\usepackage{algorithmic}

\newtheorem{lemma}{Lemma}

\newtheorem{definition}{Definition}

\newcommand{\E}{{\rm I\kern-.3em E}}

\newcommand{\R}{\rm I\!R}

\author{Konstantinos P. Tsoukatos} %, Leonidas Georgiadis}
\author{\IEEEauthorblockN{Konstantinos P. Tsoukatos} \\
\IEEEauthorblockA{Department of Computer Science and Engineering \\
Technological Education Institute of Thessaly, Greece }
}

\title
{Reciprocity--driven Sparse Network Formation}

\begin{document}

\maketitle

\begin{abstract}
A resource exchange network is considered, where exchanges among nodes are based on reciprocity.
Peers receive from the network an amount
of resources commensurate with their contribution.
%For a given connectivity graph, it is known that
%equilibrium allocations are obtained by the solution to an Eisenberg--Gale convex program.
We assume the network is fully connected, and
impose sparsity constraints on peer interactions.
%Starting with a fully connected network, we
%impose sparsity constraints on peer %interactions.
Finding the sparsest exchanges that achieve
a desired level of reciprocity is in general NP-hard.
To capture near--optimal allocations, we introduce variants of the
Eisenberg--Gale convex program with sparsity penalties. 
We derive decentralized algorithms, whereby peers 
approximately compute the sparsest allocations,
by % iterative 
reweighted $\ell_1$ minimization. 
%and greedy thresholding.
The algorithms implement new proportional-response dynamics, with nonlinear pricing.
The trade-off between sparsity and reciprocity
and the properties of graphs induced by sparse exchanges are examined. 

\end{abstract}


\section{Introduction}

The unprecedented increase in wireless traffic 
poses significant challenges for mobile operators, 
who face extensive infrastructure upgrades 
to accommodate the rising demand for data.
To ease strain on networks,
a viable alternative 
% to extensive infrastructure upgrades 
seeks to
take advantage of already deployed resources, that presently 
remain underutilized.
For example, in  device-to-device 
communications, devices in close proximity may establish either direct links, 
or indirect communication via wireless relays,
altogether bypassing the cellular infrastructure.
In recently launched Wi-Fi internet services, 
e.g., FON
(fon.com), 
Open Garden (opengarden.com), Karma  (yourkarma.com), 
sharing wireless access is a prominent feature,
and subscribers are rewarded 
% compensated 
for relaying each other's traffic.
In all these scenarios, it is important to design  mechanisms which foster cooperation and encourage user contribution, in ways that realize fair and efficient use of pooled resources.

In this paper, we study a network exchange model, where collaborative resource consumption is based on reciprocation. Incentive mechanisms based on reciprocation  have been proposed in the context of peer-to-peer and user-provided networks \cite{IGH14}, \cite{ZFP15}. Participant nodes earn credits (or virtual currency) for assisting other nodes in transmitting their data to the destination. 
Ideally, reciprocation implies that each peer receives back from the network an amount of resources or utility equal to what he contributed to other users. However, such perfect reciprocation may in general not be feasible, due to constraints arising from network structure/connectivity, and differing resource endowments possessed by nodes, also depending on their position in the network graph. Moreover, peers
can typically maintain a limited number of connections (in the popular BitTorrent peer-to-peer protocol users upload to at most four peers). Hence, it seems reasonable to explore situations where graphs representing exchanges of resources among peers are in some sense 
sparse.
%\emph{sparse}.

The exchange model considered in this paper builds upon the so-called linear Fisher market in economics
(see \cite{WuZ07}, \cite{BDX11}, \cite{ZFP15}  and references),
where each participant aims to receive as much resources as possible from the market.
In this model, the optimal resource allocations are captured by a classic convex program discovered by Eisenberg and Gale 
in 1959 \cite{EiG59}.
In the context of peer-to-peer bandwidth trading,
the authors in \cite{WuZ07} 
proposed
a simple distributed algorithm called
\emph{proportional--response},
that computes solutions to the Eisenberg--Gale program, hence also equilibrium allocations in the resource exchange model:
At every time slot, each peer 
distributes his available resource to other peers in proportion to the
resources it received from them in the previous time slot.

Here, we formulate an optimization problem that balances benefits from 
reciprocation with fixed per-link costs, therefore
induces peers to maintain only a few active connections. 
Clearly, rational peers will not engage in an exchange if costs out-weight potential benefits.
We impose sparsity penalties
on peer interactions,
to reflect the fact that peers often cannot afford
the cost of establishing and maintaining a link, the associated communication overhead, etc. fixed costs, or
are simply limited by physical constraints, such as limited range of wireless devices.
% {\bf Contribution.} 
This reciprocity versus sparsity optimization 
is solved by decentralized tit-for-tat algorithms,  whereby peers communicate bids 
for each other's resource, so as to
approximately compute the sparsest allocations,
achieving close-to-perfect reciprocation.
Our algorithms implement nonlinear pricing, and extend the proportional--response dynamics of \cite{WuZ07}  to reinforce interactions where large amounts of resources are exchanged.
Starting from a complete graph, the algorithms 
prescribe how nodes can gradually form a network of exchanges, that progressively gets sparser.
As a result, the proposed schemes suggest a network formation model where directed graphs, representing sparsity-constrained resource exchanges, are constructed.
The graphs may manifest either direct reciprocation between peers, i.e., both edges 
$(i,j)$ and $(j,i)$ are typically present in the network graph, or indirect reciprocation, in which case most of the edges do not have their reverse in the graph.
We illustrate the formation of resource exchange networks by peers who achieve almost perfect reciprocation with only a small number of connections, 
and  discuss the properties of the sparse graphs in several numerical examples.

From a mathematical standpoint, the sparse exchange algorithms are derived
by applying majorization-minimization \cite{HuL04}, \cite{SBP17} and reweighted $\ell_1$ minimization \cite{CWB08}
to a combinatorial problem, 
and optimize the trade-off between reciprocation and sparsity up to a local optimum. 
Starting from different initial conditions, different local optima arise, corresponding to different resource allocations and sparse exchange graphs.

\section{System Model and Background}
\label{sec:system}

Consider a network of $N$ peers who exchange resources over a graph $G$ describing connectivity. Exchanges take place only between peers that are neighbours in the graph $G$.
Each peer allocates spare resources to other peers,
in exchange for their resources (in the future).
There exists a single resource/commodity in the network.
Peers spend their own spare resource for acquiring resources, i.e., there is no monetary budget.
Let $a_i$ be the resource endowment of peer $i$.
Let  $x_{ij}(t)$ be the amount of resource allocated from user $j$ to user $i$ at time $t=0,1, \ldots$.
Vector $\mathbf{x}_i$
denotes the allocations of peer $i$, 
and $\mathbf{x}_{-i}$ denotes the allocations of all others.
Each peer $i$ allocates the entire 
budget $a_i$ to his neighbours, 
$a_i = \sum_{j \neq i} x_{ji}$,
and receives in return a total amount $r_i(\mathbf{x}_{-i}) := \sum_{j \neq i} x_{ij}$  of resource.
We assume peers value only the resource received
from others (not their own spare resource). That is, utility is linear in the amount of received resources  $U_i(r_i) = r_i$, and each peer $i=1, \ldots, N$, allocates resources to solve
\begin{equation*}
\emph{(PEER)} \quad \max_{\mathbf{x}_i \geq 0} \: r_i(\mathbf{x}_{-i}) \quad \mbox{subject to} \quad \sum_{j \neq i} x_{ji} = a_i.
\end{equation*}
These $N$ \emph{(PEER)} problems are intertwined, because each peer's utility depends on resources received from other peers.

\textit{Notation:} Subscript in allocation $x_{ij}$ is understood as given from $j$ to $i$, similarly $b_{ij}$ is the bid of peer $j$ for resource of peer $i$, 
and  $\mu_{ij}$ the price peer $j$ charges to $i$.

{\bf   Motivation.} 
In this paper, we consider an exchange network where the connectivity graph $G$ is complete.
Every node can, in principle, engage in exchanges with everybody else.  
However, we assume that establishing and maintaining exchange links carries a cost, so that peers tend to limit the number of their active connections.
This is often the case in practice, where peers
choose a few trading partners and avoid spreading themselves thin, so as to reduce overhead, friction etc. costs associated with exchanges. 
Limits on the number of connections may
also arise due to physical or protocol constraints (e.g., in the BitTorrent peer-to-peer protocol).
In a slightly different context, reducing transaction costs is a motivation for \emph{sparse} portfolio selection \cite{BDM09}.
% parsimony, frugal connections
Here, in a similar spirit, 
we use a penalty term that encourages peers
to form sparse connections.
Our goal is to develop a
quantitative model for dynamic formation of 
exchange networks driven by reciprocation, where the directed graphs representing exchanges are sparse.

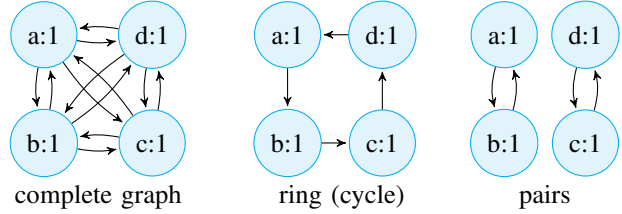
\begin{figure}
\begin{center}
\begin{tikzpicture}
	  [->,shorten >=1pt, >=stealth', auto, scale=.18, 
	  font=\small, % \sffamily
	   main node/.style={circle, draw=cyan, 
	   fill=cyan!10}]
	  \node[main node] (n1) at (0,9) {a:1};
	  \node[main node] (n2) at (0,1)  {b:1};
	  \node[main node] (n3) at (8,9)  {d:1};
	  \node[main node] (n4) at (8,1) {c:1};
	  \node[main node] (n5) at (18,9)  {a:1};
	  \node[main node] (n6) at (18,1)  {b:1};
	  \node[main node] (n7) at (25,9) {d:1};
	  \node[main node] (n8) at (25,1)  {c:1};
	  \node[main node] (n9) at (34,9)  {a:1};
	  \node[main node] (n10) at (34,1) {b:1};
	  \node[main node] (n11) at (40,9)  {d:1};
	  \node[main node] (n12) at (40,1)  {c:1};

	  \foreach \from/\to in {n1/n2,n2/n1,n1/n3,n3/n1,n1/n4,n4/n1,n2/n3,  n3/n2, n2/n4,n4/n2,n3/n4,n4/n3}
\path (\from) edge [bend right = 10] (\to);

\path	
	(n5) edge (n6)
	(n6) edge (n8)
	(n8) edge (n7)
	(n7) edge (n5)
	(n9) edge [bend right = 15](n10)
	(n10) edge [bend right =15] (n9)
	(n11) edge [bend right=15] (n12)
	(n12) edge [bend right =15] (n11);

\node at (4,-3) {complete graph};
\node at (22,-3) {ring (cycle)};
\node at (37,-3) {pairs};
	
\end{tikzpicture}
\end{center}
\caption{Exchange network with $4$ nodes. Left: Complete graph. Middle, right: Graphs with minimum number of links.}
\label{fig:4node}
\end{figure}

{\bf Example. } 
The example network of Figure 
\ref{fig:4node} illustrates the graphs implementing sparse exchanges. 
All four nodes $a,b,c,d$ have resource endowment equal to $1$. Perfect reciprocation can be realized in infinitely many ways, across the $12$ links of the complete graph (left). The sparsest exchange graphs, where each peer gives to exactly one peer $1$ unit of resource, consist of only $4$ links, arranged either in ring (middle) or pairs (right), whence we also see the sparsest solution need not be unique. In addition, the pairs configuration 
of Figure \ref{fig:4node} 
shows that the sparsest solution may partition the (initially complete) graph into disconnected components.

{\bf   Previous work.} We recap several
useful results from a large literature. Exchange network is an instance of a linear Fisher market, for which an equivalent convex formulation was given
 by Eisenberg and Gale (1959) \cite{EiG59}:
\begin{equation}
%\begin{aligned}
 \max_{\mathbf{x}} \sum_i a_i \log r_i
 \quad \text{subject to} 
  \:\: \sum_{j \neq i} x_{ji}= a_i, 
  \quad \forall i.
%\end{aligned}
\label{eq:eis59}
\end{equation}
The objective in (\ref{eq:eis59}) 
% brings to mind 
resembles the familiar proportionally-fair allocation, where each peer $i$ receives an amount of resources $r_i$ proportional to his contribution $a_i$.  
It is also similar to Kelly's \emph{NETWORK} problem \cite{Kel97}, if the contributions $a_i$ are viewed as payments. The receive vector $\mathbf{r}$ achieving optimality in (\ref{eq:eis59}) is unique, however the optimal allocation $\mathbf{x}$ is not unique, because the objective in (\ref{eq:eis59}) is \emph{not strictly} concave in $\mathbf{x}$. That is, the same optimal receive vector may be realized with different allocations. The equivalence between \textit{(PEER)}
and (\ref{eq:eis59}) can be established as follows. Let $\rho_j$  be the price at which peer $j$ 
\lq \lq sells" its resource (although no actual monetary payments mediate the exchange).
User $i$, by allocating amount $x_{ji}$ to $j$, \lq\lq purchases" back $x_{ij}= x_{ji}/\rho_j$. Hence, the total resource received by user $i$ is $r_i = \sum_{j \neq i} x_{ji}/\rho_j$. Therefore, to maximize utility in \textit{(PEER)} user $i$ allocates resources to (and consequently receives resources  $x_{ij}>0$ from) 
only peers with the largest $1/\rho_j$, i.e., the cheapest neighbors,
\begin{equation}
x_{ij} > 0 \quad  \mbox{if and only if}
\quad \rho_j = \min_{k \in {\cal N}_i} \rho_k,
\label{eq:egalloc}
\end{equation}
where ${\cal N}_i$ is the set of neighbors of $i$.
Now, to find the prices $\boldsymbol{\rho}$, consider the convex program (\ref{eq:eis59}), relax the constraints and write the Lagrangian
\[
L(\mathbf{x}, \boldsymbol{\rho}) = \sum_i a_i \log r_i
   + \sum_i \rho_i( a_i - \sum_{j \neq i} x_{ji}).
\]
The KKT conditions at the saddle point of the Lagrangian imply that either 
\[
x_{ij}=0 \quad \mbox{and}  \quad \frac{\partial L}
{\partial x_{ij}} <0 
\: \: \text{in which case} \quad
 \frac{a_i}{r_i} < \rho_j, 
\]
or 
\[
x_{ij}>0 \quad \mbox{and} \quad 
\frac{\partial L}
{\partial x_{ij}} =0 
\: \: \text{in which case} \quad
% \iff
\frac{a_i}{r_i} = \rho_j.
\]
From the equations above we deduce that  
$x_{ij}>0$  if and only if $ \rho_j = \min_{k \in {\cal N}_i} \rho_k$, which is precisely condition (\ref{eq:egalloc}). Therefore, allocations 
$\mathbf{x}$ and prices $\boldsymbol{\rho}$ solving 
\textit{(PEER)} can be computed through the Eisenberg--Gale program (\ref{eq:eis59}).

Define the \emph{exchange ratio} for each peer $i$ as the ratio of the resources $r_i$ peer $i$ receives, over the $a_i$ he allocates to others.
Since $x_{ij}= x_{ji}/\rho_j$, by summing over $i$ we get $a_j = r_j/\rho_j$, that is 
\[
\rho_j =  \frac{r_j}{a_j}, \quad \forall j.
\]
Hence, the exchange ratio coincides with the Lagrange multiplier/price $\boldsymbol{\rho}$ in the Eisenberg--Gale program (\ref{eq:eis59}). May use the term price and exchange ratio interchangeably.

When network connectivity is given,  
we summarize the following facts 
from \cite{ZFP15}, \cite{WuZ07}, \cite{BDX11},  
\cite{GIT15}:
%reveal that, 
% summarize results :
Network graph decomposes into components, and
resource exchanges take place only within each component. Since the exchange ratio is also the price (Lagrange multiplier) at which each node sells its resource, peers with high exchange ratio 
are expensive and more constrained, i.e., \lq\lq poor"
and struggle to contribute more resources.
At equilibrium, 
rational peers exchange resource only with their minimum price (cheapest/most generous) neighbours.
Exchange can be viewed as a reverse auction: Acting as sellers, peers
compete to sell their resource by lowering their prices (raising their bids), whereas, 
in the role of buyers, they purchase resource from the cheapest (highest \lq\lq bang-per-buck") neighbor.
Moreover, the prices of peers who exchange resources with each other 
are inversely proportional. That is, the price at which a node buys resources from peers is equal to the inverse of the price at which he sells his own resource.
The upshot is that, by measuring his own price, each peer can infer the price of the peers he exchange resources with.
In particular, inexpensive nodes (with prices smaller than one) 
know they interact with expensive nodes (with price larger than one). 

{\bf   New connections.} 
Previous work typically analysed exchanges in
networks where connectivity and resource endowments were a
priori fixed and immutable. Then, due to the existing connectivity and neighbouring node endowments, certain nodes may end up receiving significantly less resources than what they contribute to their peers. Unless it is possible to alter either
connectivity, or node resource endowments, rational peers with exchange ratio much lower than one have little incentive to engage in exchanges and contribute. Lack of participation
 will decrease the total amount of resources contributed to the
network, i.e., is detrimental to social welfare.

Rational nodes with low prices are motivated
to seek out new peers, who are less expensive than the ones they presently interact with.
Neighbor selection by a Gibbs sampling algorithm was proposed in \cite{ZFP15}.
By connecting with richer peers, inexpensive nodes receive more resources,
hence their exchange ratio (which is also their price)
increases. Along the way, inexpensive peers become more expensive, 
thereby also less attractive as candidates for resource exchange.
In a fully connected network, 
whenever perfect reciprocity is possible, this balancing act
may drive all exchange ratios (prices)
to one, i.e. all nodes receive from the network an amount
of resources equal to what they contribute -- perfect reciprocation.

\section{Network Formation Problems}
\label{sec:algos}

Starting from a complete network,
we allow peers to gradually form an exchange graph
that progressively gets sparser. 
First, we discuss centralized optimization problems, that 
aim to identify the sparsest interactions that guarantee a desired level of reciprocation. 
Then, we introduce variants of the Eisenberg--Gale program (\ref{eq:eis59}),
where the objective is to balance benefits from equitable allocations with fixed per-link costs, hence form only a few connections.
These formulations lead to distributed algorithms
that enable peers to compute sparse exchanges 
in a decentralized manner, by communicating bids for each other's resource. 
The algorithms are simple and natural to understand.

\subsection{Sparse Exchanges with Reciprocation Guarantees}

Let $||\mathbf{x}||_0$ denote the pseudo-norm that counts the number of nonzero entries in $\mathbf{x}$.

{\bf   Problem P0.} 
The objective is to find sparsest allocations that achieve a minimum exchange ratio at least $\theta$,
where $\theta\leq 1$:
\begin{equation}
%\begin{aligned}
 \text(P0) \quad \min_{\mathbf{x}} ||\mathbf{x}||_0
 \quad \text{s. t.} 
  \:\: \sum_{j \neq i} x_{ji}= a_i, 
 \quad \frac{r_i}{a_i} \geq 
 \theta   \quad \forall i.
%\end{aligned}
\label{eq:P0}
\end{equation}
This is a combinatorial problem (assuming
the minimum desired level $\theta$ of reciprocity is feasible), hence intractable. To find an approximate solution, 
we may replace the nonsmooth 
$||\mathbf{x}||_0$ norm by a smooth proxy. 
A typical choice, 
justified by the limit
%\begin{equation}
\[
\mathds{1}\left\{ |x| \neq 0 \right\} 
= \lim_{\epsilon \downarrow 0} 
\frac{\log(1 + |x|/\epsilon)}{\log(1+1/\epsilon)},
\]
%\end{equation}
is given by the logarithmic approximation
$\ell (\mathbf{x}) := \sum_{i,j} \log(\epsilon +x_{i,j})$, leading to the problem
\begin{equation}
%\begin{aligned}
 \quad \quad \min_{\mathbf{x}} \: \ell(\mathbf{x})
 \quad \text{s. t.} 
  \:\: \sum_{j \neq i} x_{ji}= a_i, 
 \quad \frac{r_i}{a_i} \geq 
 \theta   \quad \forall i.
%\end{aligned}
\label{eq:concave}
\end{equation}
Hence, the combinatorial objective in
(\ref{eq:P0}) has been substituted by a minimization of a concave function (\ref{eq:concave}). This is 
again hard, and can be tackled as follows.

{\bf   Problem P1.}
Successively minimize a linear upper bound to the logarithm in (\ref{eq:concave}), given by
\begin{equation}
 \quad \min_{\mathbf{x}} 
 \: \mathbf{x}^T \nabla  \ell(\mathbf{x}(t))
 \quad \text{s. t.} 
  \:\: \sum_{j \neq i} x_{ji}= a_i, 
 \quad \frac{r_i}{a_i} \geq 
 \theta   \quad \forall i,
\label{eq:LP}
\end{equation}
formed around the previous solution $\mathbf{x}(t)$, for $t = 0,1, \ldots$. This amounts to solving a series of linear programs to find a local minimum, hence can be computed efficiently. Moreover, to avoid getting trapped in local minima, a small random perturbation may be employed.

% Is it true that as epsilon goes to 0 obtain optimal solution?
%
%
{\bf   Problem P2.}
Instead of bounding the logarithm in (\ref{eq:concave}) with a linear upper bound, 
we bound with a quadratic. More
specifically, it holds that
$
\log(\epsilon + x) \leq q(x,\bar{x})+k,
$
where
\begin{equation}
q(x, \bar{x}) :=
\left\{\begin{array}{ll}
 x^2/(2 \delta (\epsilon + \delta)),
&
0 \leq x \leq \delta \\
x^2/(2 \bar{x} (\epsilon + \bar{x})),
&  
x > \delta,
 \end{array} \right.
 \label{eq:logquad}
\end{equation}
for appropriate constant $k$ and small $\delta >0$. Starting from an iterate $\mathbf{x}(t)$,
we use  (\ref{eq:logquad}) to bound (\ref{eq:concave})
and get a quadratic program in $\mathbf{x}$:
\begin{equation}
%\begin{aligned}
  \min_{\mathbf{x}}
 \sum_{i,j} q(x_{ij}, x_{ij}(t))
 \:\: \text{s. t.} 
  \:\: \sum_{j \neq i} x_{ji}= a_i, 
 \:\frac{r_i}{a_i} \geq 
 \theta   \quad \forall i.
%\end{aligned}
\label{eq:P2}
\end{equation}
Problem (\ref{eq:P2}) subsequently reduces to a linear system computing $2N$ multipliers from $2 N$ linear equations.
As in (\ref{eq:LP}), we apply the majorization-minimization procedure
\cite{CWB08}, hence
solve a sequence of quadratic programs to obtain the final solution (IRLS algorithm).
Instead of solving each quadratic program completely, we may run one (or a few) iterations 
towards solution (e.g. fixed-point iteration) of linear system. We anchor a new upper bound to the computed allocations, and resume iterations for the updated linear equations.

The solutions discussed above are centralized. 
In the following, we focus on
distributed algorithms, obtained by
balancing reciprocation with  
a penalty that encourages sparse exchanges.

\subsection{Eisenberg--Gale Program with Sparsity Penalty }

We consider an  Eisenberg--Gale program (\ref{eq:eis59}) augmented with a sparsity promoting term

\begin{equation}
\begin{aligned}
\text(EG) \quad \max_{\mathbf{x}} \: \sum_i a_i \log r_i
- c \sum_{i,j} \log( \epsilon + x_{ij} ) \\
 \text{subject to} \quad \sum_{j \neq i} x_{ji} = a_i,    
 \quad  \forall i.
\end{aligned}
\label{eq:EG}
\end{equation}
Optimization (\ref{eq:EG}) is nonconvex because the objective is a difference of two concave functions.
We will derive a distributed algorithm that computes an approximate solution.  First, relax the constraints, introduce the multipliers 
$\boldsymbol{\lambda}$
and write the Lagrangian
\[
L(\mathbf{x}, \boldsymbol{\lambda}) = \sum_i a_i \log r_i
- c \sum_{i,j} \log( \epsilon + x_{ij} )
   + \sum_i \lambda_i ( a_i - \sum_{j \neq i} x_{ji} ).
\]
The dual optimization requires solving
the relaxed primal
\begin{equation}
\max_{\mathbf{x}} L(\mathbf{x}, \boldsymbol{\lambda}).
\label{eq:L}
\end{equation}
A local maximum for (\ref{eq:L}) can be determined in a iterative fashion using majorization-minorization.
% procedure \cite{SBP17}. 
To that end, we bound the 
logarithms $\log (\epsilon + x_{ij})$
using
\begin{equation}
\log y \leq \log y_0 + \frac{y-y_0}{y_0}.
\label{eq:logbd}
\end{equation}
Next, with a logarithmic change of variables
${\tilde x}_{ij} := \log x_{ij}$, define
the function
%\begin{equation}
%\label{eq:logsumexp}
\[
\phi(\tilde{\mathbf{x}}) : =
\sum_{i} a_i \log ( \sum_{j \neq i} 
e^{\tilde x_{ij}} ), 
\]
%\end{equation}
so that $ \phi(\tilde{\mathbf{x}}) =
\sum_i a_i \log r_i(\mathbf{x}).
$
The convexity of the log-sum-exp function 
\cite[page 74]{boyd} implies that $\phi$ is convex in $\tilde{ \mathbf{x}}$, therefore it holds that
\begin{equation}
\phi(\tilde{\mathbf{x}}) \geq \phi(\tilde{\mathbf{x}}(t)) +  
(\tilde{\mathbf{x}} - \tilde{\mathbf{x}}(t))^T 
\nabla \phi(\tilde{\mathbf{x}}(t)). 
\label{eq:phibound}
\end{equation}
Making use of inequalities 
(\ref{eq:logbd}) and 
(\ref{eq:phibound}),
we lower bound the Lagrangian $L$
by a surrogate function $g$
anchored at $\tilde{\mathbf{x}}(t)$,
\[
L(\tilde{\mathbf{x}}, \boldsymbol{\lambda})
\geq g(\tilde{\mathbf{x}} | 
\tilde{\mathbf{x}}(t)),
\]
constructed as
\begin{eqnarray}
g(\tilde{\mathbf{x}} | \tilde{\mathbf{x}}(t)) & := &
\phi(\tilde{\mathbf{x}}(t)) +  
(\tilde{\mathbf{x}} - \tilde{\mathbf{x}}(t))^T 
\nabla \phi(\tilde{\mathbf{x}}(t))
\nonumber \\
& &
- c \sum_{i,j} ( \log ( \epsilon + e ^{\tilde{x}_{i,j}(t)} ) - 1) 
\nonumber \\ 
& & - \sum_{j} 
\left( \lambda_j + w(e^{\tilde{x}_{ij}(t)} ) \right )
 \: e^{\tilde{x}_{ij}}.
\label{eq:surg}
\end{eqnarray}
The weights $w$ above are defined as in \cite{CWB08} by  
\begin{equation}
\label{eq:w}
w(x) := \frac{c}{\epsilon + x}.
\end{equation}
Function $g(\tilde{\mathbf{x}} | \tilde{\mathbf{x}}(t))$ in (\ref{eq:surg}) is concave in the transformed allocations $\tilde{\mathbf{x}}$ 
and easy to maximize. Successive maximization of the lower bound (\ref{eq:surg}) yields the iteration 
\[
\tilde{\mathbf{x}}(t+1) = \arg \max_{\tilde{\mathbf{x}}}
g (\tilde{\mathbf{x}} | \tilde{\mathbf{x}}(t) ), 
\]
where, at each time $t=0,1,\ldots$, multipliers $\boldsymbol{\lambda}$ are also updated to satisfy the node endowment constraints in (\ref{eq:EG}). We differentiate (\ref{eq:surg}) with respect to $\tilde{\mathbf{x}}$,
transform back to the $\mathbf{x}$ domain,
and, after some algebra, arrive at the following solution:

Each peer $i$ communicates at time $t=0,1,\ldots$  its exchange ratio
\begin{equation}
\rho_{i}(t) :=  \frac{r_i(t)}{a_i} 
\label{eq:rhoEG}
\end{equation}
to other peers.

Let the bids of peer $j$ for peer $i$'s resource be
\begin{equation}
b_{ij}(\mathbf{x}(t), \lambda_i) :=  \frac{x_{ji}(t)}{\rho_j(t)} \frac{1}{\lambda_i + w(x_{ji}(t))},
\label{eq:bidEG}
\end{equation}
where the weights $w$ are defined in (\ref{eq:w})
and $\lambda_i$ is the multiplier associated with the budget constraint for peer $i$. 
In market terms, $b_{ij}$ is the amount of resource peer $i$ can purchase from $j$ at price (per unit)
$\rho_j(t) (\lambda_i + w(x_{ji}(t))) $ by paying $x_{ji}(t)$. Note that pricing is nonlinear; 
price per unit decreases as payment $x_{ji}(t)$ increases, and asymptotically drops to $\lambda_i \rho_j(t)$ as payment $x_{ji}(t)$ goes to infinity.

Each peer $i$ selects  $\lambda_i$ to exhaust the entire budget
\begin{equation}
a_i = \sum_{j \neq i} b_{ij}(\mathbf{x}(t), \lambda_i),
\label{eq:blambda}
\end{equation}
this can be computed by bisection search. 
Check that if endowment $a_i$ of peer $i$ is large, then, all other quantities in (\ref{eq:blambda}) remaining fixed, multiplier 
$\lambda_i$ (which is also the price associated
with $i$'s endowment) will be small, i.e., peer $i$ will be less resource constrained, as expected.

Finally, peer $i$ allocates resources to $j$ proportionally to bids
\begin{equation}
x_{ji}(t+1) = b_{ij}(\mathbf{x}(t), \lambda_i)
\label{eq:xEGSParse}
\end{equation}
where the bids $b_{ij}$ are defined by (\ref{eq:bidEG}) and multipliers $\lambda_i$ solve (\ref{eq:blambda}).
This is a proportional-response with  nonlinear price discrimination.
We call the algorithm an EG-sparse proportional-response (EGsPaRse). 

\begin{algorithm}[H]
\caption{ Eisenberg-Gale Sparse Proportional Response  \textsc{(EGsPaRse)}}
\begin{algorithmic}[1]
% \REQUIRE in
% \ENSURE  out
% \\ \textit{Initialisation t=0 }: 
%  \STATE first statement
% \\ \textit{LOOP Process}
 \STATE Initialization (time $t=0$):
  Peers allocate resources $\mathbf{x}(0)$ either equally or randomly. 
 \REPEAT
%  \FOR {$i = l-2$ to $0$}
  \STATE Each peer $i$ computes its exchange ratio $\rho_i(t)$ (\ref{eq:rhoEG}) and communicates it 
  to the network.
  \STATE Each peer $i$ determines the bids $b_{ij}(t)$ for its resource from (\ref{eq:bidEG}), 
  where the multiplier  $\lambda_i$ solves (\ref{eq:blambda}).
  \STATE Each peer $i$ allocates resources $x_{ji}(t+1)$ according to (\ref{eq:xEGSParse})
%  \IF {($i \ne 0$)}

%  \ENDIF
%  \ENDFOR
% \RETURN $P$
 \STATE $ t \gets t +1 $
 \UNTIL Convergence
\end{algorithmic}
\end{algorithm}

When there is no sparsity-promoting penalty $c = 0$, the recursion becomes
\begin{equation}
\label{eq:PR-twostage}
% x_{ij}^{(t+1)}
x_{ij}(t+1) = a_j  \: x_{ij}(t) \: 
\frac{a_i}{r_i(t)} \bigg / \sum_{k \neq j} x_{kj}(t) \frac{a_k}{r_k(t)}.
\end{equation}
Updates (\ref{eq:PR-twostage}) coincide with the standard proportional-response dynamics of \cite{WuZ07}.
To verify this, observe that iteration (\ref{eq:PR-twostage}) corresponds to two steps
of proportional-response: In the numerator, peer $i$ reciprocates $j$ by charging
a constant per-unit price $r_i(t)/a_i$ (linear pricing), likewise each peer $k$ in the denominator, and for peer $j$ reciprocating $i$ 
in the entire fraction.

In the general case $c >0$, 
peers are required to
communicate either their exchange ratio, or the multiplier $\lambda$ (which is also related to the exchange ratio).  This
implicitly assumes peers declare their true ratio. In practice, peers may be unwilling to disclose their exchange ratio (due e.g. to privacy) or strategically misreport it, to extract additional resources. Such strategic/non-cooperative behaviour by peers who anticipate the effect of reporting their ratio may result in loss of optimality.

\subsection{ An Alternative Formulation: SPaRse Algorithm}
 We next turn to an alternative formulation, which leads to an intuitively appealing algorithm. Recall the definition of the Kullback--Leibler divergence
between two vectors,
\[
D(\mathbf{u}, \mathbf{v}) := \sum_i u_i \log \frac{u_i}{v_i} - \sum_i (u_i - v_i),
\quad \mathbf{u}, \mathbf{v}\geq 0.
\]
Inspection of the Eisenberg--Gale program  (\ref{eq:eis59}) shows it is 
equivalent to minimizing the divergence 
$D(\mathbf{a}, \mathbf{r})$ between allocated 
$\mathbf{a}$ and received $\mathbf{r}$ resources (subject to constraints).
It is natural to wonder whether we may seek to minimize $D(\mathbf{r}, \mathbf{a})$ instead of $D(\mathbf{a}, \mathbf{r})$; 
although divergence is in general not symmetric.
It turns out that the former optimization 
also captures the optimal allocations,
a result due to Shmyrev \cite{Smy09} 
(see also discussion in \cite{BDX11}). 
The key advantage of this alternative formulation is that it nicely fits the proportional-response dynamics.
Hence, we consider a
convex program equivalent to (\ref{eq:eis59}),
obtained from $\min_\mathbf{x}D(\mathbf{r}, \mathbf{a})$, together with a sparsity penalty:

\begin{equation}
\begin{aligned}
\text(S) \quad \min_{\mathbf{x}} \: D(\mathbf{r}, \mathbf{a})
+ c \sum_{i,j} \log( \epsilon + x_{ij} ) \\
 \text{subject to} \quad \sum_{j \neq i} x_{ji} = a_i,    
 \quad  \forall i.
\end{aligned}
\label{eq:S}
\end{equation}
Optimization problem (\ref{eq:S}) is nonconvex; 
we will derive an algorithm that computes a local minimum using the minorization-majorization procedure \cite{SBP17}. 
%First, we form an upper bound to the objective %function, then linearize log, add a K-L %divergence
%and then do 
The updates can be expressed in terms of a {\em Bregman divergence} $B_{h}$, %(\mathbf{u}, \mathbf{v})$
associated with the convex negative entropy  function $h$.

\begin{definition}
Let $\psi: X \rightarrow \R $ be a strongly convex 
function on a convex
set $X$. The Bregman divergence $B_{\psi}: X \times X \rightarrow \R$ associated
with the strongly convex function $\psi$ is defined by
\begin{equation}
B_{\psi}(\mathbf{u}, \mathbf{v}):=  \psi(\mathbf{u}) - \psi(\mathbf{v}) -
(\mathbf{u} - \mathbf{v})^T \nabla \psi(\mathbf{v}).
% |_{\mathbf{p}=\mathbf{p}^r}
\label{eq:bregmandef}
\end{equation}
\end{definition}
The Bregman divergence is a distance--like function, as it satisfies $B_{\psi}(\mathbf{u}, \mathbf{v}) \geq 0$ for all $\mathbf{u}, \mathbf{v}$, thanks to the convexity of $\psi$.
For example, $\psi(\mathbf{u}) =  \frac{1}{2}|| \mathbf{u}||^2$ induces the usual Euclidean distance
$B_\psi(\mathbf{u}, \mathbf{v}) = \frac{1}{2} || \mathbf{u}- \mathbf{v}||^2$,
however the Bregman divergence is in general not symmetric (for more properties see e.g. \cite{BeT03}).
The Bregman distance generated by the negative entropy  
\begin{equation}
h(\mathbf{r}) :=  \sum_i r_i \log r_i, \quad \mathbf{r} \geq 0,
\label{eq:entropy}
\end{equation}
is the Kullback--Leibler divergence:
\begin{equation}
B_{h}(\mathbf{u}, \mathbf{v}) = D(\mathbf{u}, \mathbf{v}),
\quad \mathbf{u}, \mathbf{v}\geq 0. 
\label{eq:BhD}
\end{equation}
This particular choice of Bregman function (instead of usual Euclidean distance) is motivated by the fact that entropy better reflects the geometry of the simplex constraints  \cite{BeT03},  \cite{BDX11} (so that the latter are easily eliminated).
%(itakura-saito is generated by -log(y))

Let $f(\mathbf{x})$ be the  objective function in (\ref{eq:S}). 
In the majorization step, a point 
$\mathbf{x}(t)$ is used to anchor a surrogate function
$g(\mathbf{x} | \mathbf{x}(t))$
which upper bounds $f$, 
\[
f(\mathbf{x}) \leq g(\mathbf{x} | \mathbf{x}(t)), \quad \mathbf{x} \geq 0,
\]
and is easy to minimize. Function $g$ is chosen to be tight at $\mathbf{x}(t)$, i.e.,  
$f(\mathbf{x}(t)) = g(\mathbf{x}(t) | \mathbf{x}(t))$. 
In the minorization step, the upper bound is minimized with respect to $\mathbf{x}$,  generating a sequence  
\begin{equation}
\mathbf{x}(t+1) = \arg \min_{\mathbf{x}}
g(\mathbf{x} | \mathbf{x}(t)), 
\label{eq:mmupdate}
\end{equation}
for each $t=0,1,\ldots.$
We form the surrogate $g(\mathbf{x} | \mathbf{x}(t))$ as follows:  
Write the divergence 
$D(\mathbf{r}, \mathbf{a})$ as
\begin{equation}
D(\mathbf{r}, \mathbf{a}) = h(\mathbf{r})
- \sum_i r_i \log a_i - (r_i - a_i).
\label{eq:Dra}
\end{equation}
 Because of (\ref{eq:BhD}), negative entropy (\ref{eq:entropy}) satisfies
\begin{equation}
h(\mathbf{r}) = h(\mathbf{r}(t)) + 
(\mathbf{r}-\mathbf{r}(t))^T 
\nabla h (\mathbf{r}(t)) + D(\mathbf{r}, \mathbf{r}(t)).
\label{eq:hD}
\end{equation}
The divergence $D(\mathbf{r}, \mathbf{r}(t))$ in (\ref{eq:hD}) is bounded
using Lemma \ref{lem:Dx}
(end of Section \ref{sec:algos}).
The logarithm in (\ref{eq:S}) is bounded
by the first-order Taylor expansion
%\begin{equation}
%f(\mathbf{x}) = h(\mathbf{r})  - \sum_i r_i \log a_i - \sum (r_i - a_i) + I_B + I_C
%
%\label{eq:objS}
%\end{equation}
\begin{equation}
\log y \leq \log y_0 + \frac{y-y_0}{y_0},
\label{eq:logbound}
\end{equation}
as is customary in the reweighted $\ell_1$ minimization \cite{CWB08} framework.
Inserting (\ref{eq:Dra}) and (\ref{eq:hD}) in the objective (\ref{eq:S}) and taking into account inequalities (\ref{eq:logbound}) and (\ref{eq:Drx})
gives
\begin{eqnarray}
g(\mathbf{x} | \mathbf{x}(t)) & = &
D(\mathbf{x}, \mathbf{x}(t))
+
h(\mathbf{r}(t)) + 
(\mathbf{r}-\mathbf{r}(t))^T 
\nabla h (\mathbf{r}(t)) \nonumber \\
& &
+ c \sum_{i,j} 
\log (\epsilon + x_{ij}(t))
+ \frac{x_{ij}-x_{ij}(t)}{\epsilon + x_{ij}(t)} \nonumber \\
& & -
\sum_i r_i \log a_i - (r_i - a_i).
\label{eq:g}
\end{eqnarray}
The updated allocations $\mathbf{x}(t+1)$ 
are computed by minimizing the surrogate (\ref{eq:g})
in (\ref{eq:mmupdate}). 
After some algebra,
also making use of 
$
\partial D(\mathbf{x}, \mathbf{x}(t)) /
\partial x_{ij} = \log x_{ij} - \log x_{ij}(t)$,
we get 
\begin{equation}
x_{ij}(t+1) = x_{ij}(t) \frac{a_i}{r_i(t)}
\exp \left(- \frac{c}{\epsilon + x_{ij}(t)}\right). 
\label{eq:xSmin}
\end{equation}
Finally, allocations (\ref{eq:xSmin}) are normalized to satisfy the endowment 
constraint $\sum_{i \neq j} x_{ij} = a_j$ 
for each peer $j$.
We thus arrive at a second algorithm for sparse proportional-response, where nonlinear prices (with an exponential factor) are charged to users:

Each peer $i$ computes the price $\mu_{ji}(t)$ (per unit resource) charged to peer $j \neq i$ at time $t=0,1,\ldots$ as
\begin{equation}
\mu_{ji}(t) :=  \frac{r_i(t)}{a_i} \: \exp \left(  \frac{c}{\epsilon + x_{ij}(t)} \right).
\label{eq:muS}
\end{equation}
Pricing above is nonlinear, because $\mu_{ji}(t)$ 
depends on the amount of resource $x_{ij}(t)$ (payment) offered to $i$,
inside the exponential. 
The higher the resource $x_{ij}(t)$  (payment) offered by peer $j$ to peer $i$, the lower the price 
$\mu_{ij}(t)$  (per unit resource) charged to $i$, and the price converges to the exchange 
ratio $r_i(t)/a_i$ as payment $x_{ij}(t)$ goes to infinity.
Hence, the proposed dynamics reinforce exchanges that involve large amounts of resources.

Next, each peer $i$ communicates to other peers $j$ the price $\mu_{ji}(t)$. 
Peer $i$ computes the bids of other peers $j$ for
$i$'s resource as
\begin{equation}
b_{ij}(\mathbf{x}(t)) :=  \frac{x_{ji}(t)}{\mu_{ij}(t)},
\quad j \neq i.
\label{eq:bidS}
\end{equation}
This corresponds to the amount of resource with which $j$ intends to reciprocate $i$. Alternatively, peers can communicate directly the bids instead of prices.
Bid $b_{ij}$ is also the number of resource units that peer $i$ can purchase from $j$ with total payment $x_{ji}(t)$, at price $\mu_{ij}(t)$.  

Subsequently, peer $i$ allocates his resource to peer $j$ proportionally to the received bids, 
\begin{equation}
x_{ji}(t+1) = a_i \: \frac{ b_{ij}(\mathbf{x}(t)) }{\displaystyle \sum_{k \neq i} b_{ik}(\mathbf{x}(t))},
\quad j \neq i.
\label{eq:xSparse}
\end{equation}
This is a proportional-response with nonlinear price discrimination,
where larger amounts of resource are \lq \lq sold" 
at lower per-unit price (discount).
We call this algorithm a Shmyrev-sparse proportional-response (SPaRse).

% wholesale vs retail

\begin{algorithm}[H]
\caption{ Shmyrev Sparse Proportional Response        
\textsc{(SPaRse)}}
\begin{algorithmic}[1]
% \REQUIRE in
% \ENSURE  out
% \\ \textit{Initialisation t=0 }: 
%  \STATE first statement
% \\ \textit{LOOP Process}
 \STATE Initialization (time $t=0$):
  Peers allocate resources $\mathbf{x}(0)$ either equally or randomly. 
 \REPEAT
%  \FOR {$i = l-2$ to $0$}
  \STATE Each peer $i$ computes the nonlinear price $\mu_{ji}(t)$ (\ref{eq:muS}) (per unit resource) he charges to each peer $j$, and communicates $\mu_{ji}(t)$ to peer $j$.
  \STATE Each peer $i$ determines the bids $b_{ij}(t)$ for its resource by peer $j$ using (\ref{eq:bidS}). 
  \STATE Each peer $i$ allocates resources $x_{ji}(t+1)$ proportionally to bids (\ref{eq:xSparse}).
%  \IF {($i \ne 0$)}

%  \ENDIF
%  \ENDFOR
% \RETURN $P$
 \STATE $ t \gets t +1 $
 \UNTIL Convergence
\end{algorithmic}
\end{algorithm}
Each round of SPaRse has ${\cal O}( N^2 )$ computation and communication complexity. If there is no sparsity penalty (set $c=0$) we recover again recursion (\ref{eq:PR-twostage}), which is
the standard proportional-response \cite{WuZ07}.
As will be seen in the numerical results of Section \ref{sec:numerical}, the variant with Equal first round allocation $\mathbf{x}(0)$ tends to generate graphs with 
mostly \emph{direct} reciprocation,
while Random first round leads to \emph{indirect} reciprocation,  This is likely due to the fact that a random initial allocation adds uncertainty
and erases symmetry, so that it gets impossible to recover the more orderly direct reciprocation.

The analysis above can be extended to
address a slightly different model, 
where each peer is constrained
by the maximum number of active connections
it can maintain at all time slots.

\begin{lemma} 
\label{lem:Dx}
For all $\mathbf{x}, \mathbf{y} \geq 0$ 
it holds that
\begin{equation}
\label{eq:Drx}
D(\mathbf{r}(\mathbf{x}), \mathbf{r}(\mathbf{y}))
\leq D(\mathbf{x}, \mathbf{y}).
\end{equation}
\end{lemma}
\begin{IEEEproof}
Inequality  (\ref{eq:Drx}) follows from the 
joint convexity of the function $d(x,y) = x \log (x/y)$ in $(x,y)$ and Jensen's inequality
\cite{BDX11}.
\end{IEEEproof}

\begin{figure}[t]
\centering
\includegraphics[width=8.5cm]{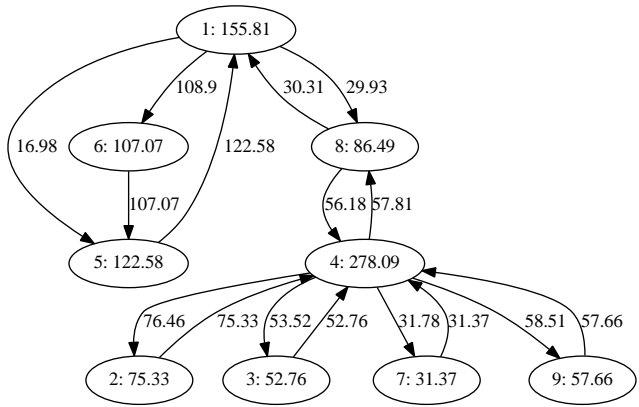}
\caption{SPaRse-Equal: Exchange graph in a $N=9$ node network, $T=10,000$ iterations. }
\label{fig:net9}
\end{figure}

\section{Numerical Results}
\label{sec:numerical}

We evaluate the performance of the
SPaRse proportional-response algorithm in several numerical examples; EGsPaRse 
is omitted for brevity.
% since it seems less suitable for implementation.
The examples showcase the formation of sparse exchange graphs by peers who communicate bids/prices in a distributed manner, and compute allocations that achieve close to perfect reciprocation (minimum exchange ratio near one).
We discuss the influence of the initial split
(SPaRse-Equal versus SPaRse-Random variants)
on the properties of the induced graphs in terms of direct/indirect reciprocation, the role of the link cost parameter $c$, and the temporal effects (number of iterations $T$) on the sparsity and fairness of the resulting allocations. 

The SPaRse algorithm is applied to a $9$-node network, where node endowments are shown in the circles, i.e., node $1$ endowment is $155.81$.
After $10,000$ iterations of SPaRse-Equal (with $c=0.1$, $\epsilon=0.01$), a graph with $16$ links is generated, shown in Figure \ref{fig:net9}, together with the computed  allocations $x_{ij}$. The minimum exchange ratio is $0.981$, and the divergence 
between received and allocated resources is 
$D(\mathbf{r},\mathbf{a})=0.125$.
We see the majority of links are bidirectional: Only links  $1 \rightarrow 6$, and $6 \rightarrow 5$ do not have their reverse in the graph, so among these three nodes                      
 \emph{indirect} reciprocation takes place.

\begin{figure}[t]
\centering
\includegraphics[width=9cm]{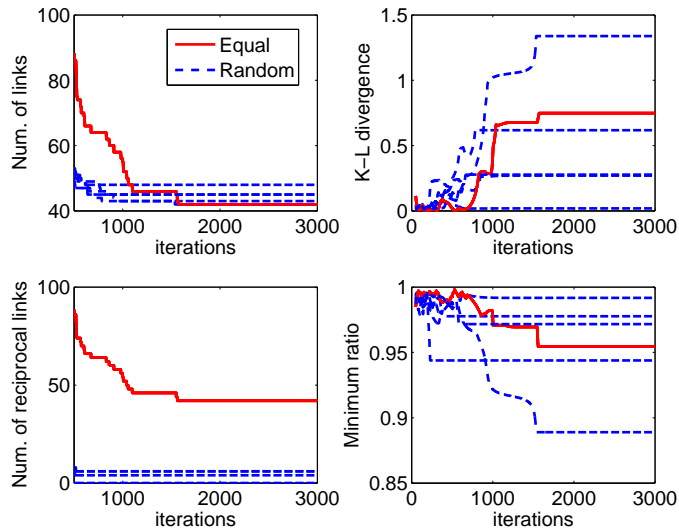}
\caption{SPaRse convergence in a $N=25$ node network: Equal (solid) vs. Random (dashed) initial allocation.}
\label{fig:N25}
\end{figure}
We next consider a $25$-node network, with sample mean node endowment $\bar{a}=106.22$ 
and standard deviation $std(a)=53.48$, drawn from a lognormal $\log a_i \sim {\cal N}(4.5, 0.25)$ distribution.
We compare influence on the resulting allocations of the Random and Equal initial splits.
Figure \ref{fig:N25} shows six sample paths
of the SPaRse algorithm (with $c=0.2$,
$\epsilon=0.01$); solid red line corresponds to Equal initial split of resources, and dashed blue lines correspond to five 
Random initial splits.  It appears that 
starting with Equal allocation requires more time to converge.
We see that, in general, convergence takes place to different allocations, and slightly different minimum exchange ratios, which are larger than $0.9$, not too far from $1$. 
The top left plot shows
that the cardinality of the final allocation is roughly the same under both Random and Equal, i.e., regardless of initial conditions. More interestingly, the bottom left plot in Figure \ref{fig:N25} suggests  qualitatively different behavior of the two variants: (a) SPaRse-Equal forms a graph that implements \emph{direct} reciprocation (number of reciprocal links is almost equal to total number of links, at about $45$); while (b) SPaRse-Random 
generates graphs that implement
\emph{indirect} reciprocation, as there are very few reciprocal links.

\begin{figure}[t]
\centering
\includegraphics[width=9cm]{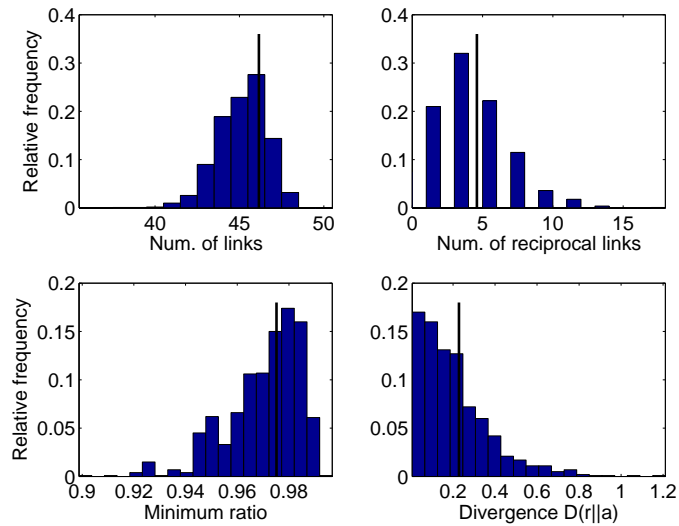}
\caption{SPaRse-Random in $N=25$ node network.  Histogram of metrics over 1000 runs with random initial splits. }
\label{fig:N25hist}
\end{figure}

The impact of different random initial allocations is quantified in the the same $25$-node network, and node endowments as Figure \ref{fig:N25}.
We run SPaRse-Random $1,000$ times (with $c=0.1$, $\epsilon = 0.01$), each time with a different random split of resources in the first round.
Runs are $5,000$ iterations long, by then allocations have converged. 
We record four performance metrics: (i) the cardinality of the final allocation (the number of directional links in the resulting exchange network), (ii) the reciprocity (i.e., the number of links for which their reciprocal is also in the graph), 
(iii) the minimum exchange ratio over the $25$ nodes,
and (iv) the 
divergence $D(\mathbf{r}, \mathbf{a})$  between received $\mathbf{r}$ and allocated $\mathbf{a}$ resources. Figure \ref{fig:N25hist} shows histograms and mean values  (vertical black line) for all $4$ metrics. 
%It is clear that
% different initializations may lead to different % local optima. 
We see that 
SPaRse-Random usually achieves a
minimum exchange ratio larger than $0.92$,  with sparse graphs consisting of less than $50$ edges, out of $25 \times 24 = 600$ totally in a complete graph with $25$ nodes.
The top histograms (cardinality of $\mathbf{x}$, reciprocity) 
once more indicate that graphs generated by SPaRse-Random manifest mostly indirect reciprocation,
since (on the average) only about $4$ out of the $46$ links are reciprocal.

\begin{figure}[t]
\centering
\includegraphics[width=8.5cm, height=3.5cm]{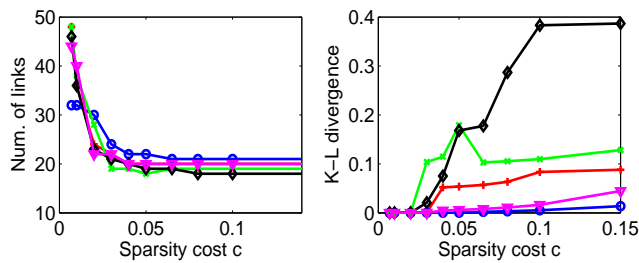}
\caption{SPaRse-Equal: Performance under different sparsity parameters $c$ ($N=11$ node network). }
\label{fig:CardDiv}
\end{figure}

The role
of sparsity parameter $c$ is examined in Figure \ref{fig:CardDiv}.
In a $N=11$ node network, five endowment vectors are randomly drawn from
the same lognormal distribution as before.
For each endowment vector we run SPaRse-Equal with different sparsity parameters $c$ and record the cardinality of the resulting allocation, and the divergence $D(\mathbf{r}, \mathbf{a})$, to
get five cardinality and divergence curves.
The duration of each run is $10^4$ iterations. 
As $c$ decreases, the algorithm computes more fair allocations (smaller divergence, larger minimum exchange ratio), but takes longer to converge. 
Decreasing $c$ below $0.05$ (while average endowment is about $100$) yields close to zero divergence, i.e., perfect reciprocation (Figure \ref{fig:CardDiv}, right). However, for $c$ smaller than $0.05$, and
when computations stop after $10^4$ iterations, we see that almost zero divergence 
is accompanied by an increase in 
number of links in the graph  (Figure \ref{fig:CardDiv}, left).

\begin{figure}[t]
\centering
\includegraphics[width=8.5cm]{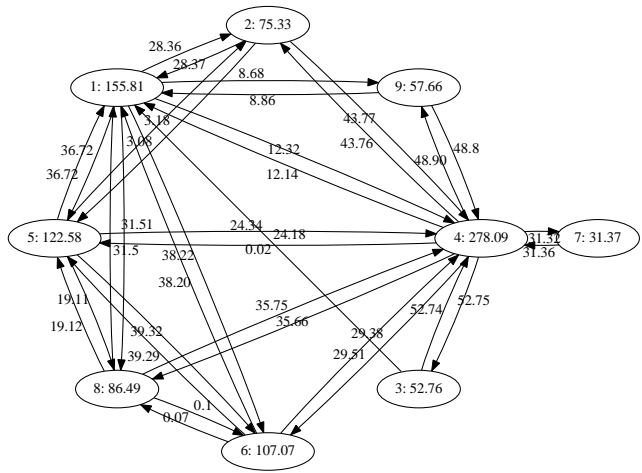}
\caption{SPaRse-Equal: Exchange graph in a $N=9$ node network, $T=500$  iterations. }
\label{fig:net9T500}
\end{figure}

The discussion above suggests that, by tuning the parameters $c$ and $\epsilon$, our model can generate graphs with various levels of sparsity and reciprocation, which also evolve temporally 
as the allocation of resources changes over the course of time. 
Apart from the static graphs that arise after SPaRse converges, one may also take a snapshot of the network at some \emph{finite} time, during the transient. 
For example, at time $t=0$ let us start with the $9$-node endowments of Figure \ref{fig:net9} 
and apply SPaRse in a complete graph, which sparsifies as time elapses. By stopping early after $500$ iterations, we obtain the graph shown in Figure \ref{fig:net9T500}. This consists of $35$ links (as compared to $16$ links in Figure          \ref{fig:net9}),
where only link $3\rightarrow 1$ is not directly reciprocated (but has small allocation $0.02$). The minimum exchange ratio is $0.998$, and divergence $D(\mathbf{r},\mathbf{a})=0.001$, while 
the respective values in Figure \ref{fig:net9} were      $0.981$ and $0.125$. Therefore, graph in Figure \ref{fig:net9T500} is less sparse than  Figure \ref{fig:net9}, but realizes more fair exchanges. 
A common feature of the allocations 
in both Figures \ref{fig:net9} and \ref{fig:net9T500}
is that
low endowment nodes apparently never exchange resources with each other.

\section{Conclusion}

We studied a resource exchange network where exchanges among nodes are based on reciprocity.
To incorporate costs of establishing and maintaining active connections, we imposed sparsity penalties on peer interactions.
Finding the sparsest graphs that achieve a certain level of reciprocation is in general NP-hard.
We proposed decentralized algorithms,  
that enable peers to approximately compute the sparsest allocations, 
by generalized proportional-response dynamics, with nonlinear pricing.
Numerical results illustrate the performance of the SPaRse algorithms and the formation of exchange graphs by peers who achieve
close-to-perfect reciprocation (minimum exchange ratio near one), 
%while maintaining
%a limited number of connections.
in a network with a limited number of
active connections.

\bibliographystyle{ieeetr}

\bibliography{p2pbib}

\end{document}